\documentstyle[twocolumn,psfrag,epsfig,prl,aps,amsmath,amssymb,float]{revtex}
\begin{document}
\newfloat{figure}{ht}{aux}
\draft
\twocolumn[\hsize\textwidth\columnwidth\hsize\csname
@twocolumnfalse\endcsname
\title{Self-adapting method for the localization of quantum critical points using Quantum Monte Carlo techniques}
\author{F. Alet and E. S. S\o rensen}
\address{Laboratoire de Physique Quantique \& 
UMR CNRS 5626, Universit\'e Paul Sabatier, 31062 Toulouse, France}  
\date{\today}
\maketitle
\begin{abstract}
A generalization to the quantum case of a recently introduced algorithm
(Y.~Tomita and Y.~Okabe, Phys. Rev. Lett. {\bf 86}, 572 (2001)) for the
determination of the critical temperature of classical spin models is
proposed. We describe a simple method to automatically locate critical points
in (Quantum) Monte Carlo simulations. The algorithm assumes the
existence of a finite correlation length in at least one of the two
phases surrounding the quantum critical point. We
illustrate these ideas on the example of the critical
inter-chain coupling for which coupled antiferromagnetic
$S=1$ spin chains order at $T=0$.
Finite-size scaling relations are used to determine the exponents,
$\nu=0.72(2)$ and $\eta=0.038(3)$ in agreement with previous estimates. 
\end{abstract}

\vskip2pc]
%============================================================================
% BODY OF PAPER

% \section{Introduction}

For studies of phenomena related to phase transitions, either classical
or quantum, a precise determination of the critical point is of crucial
importance, especially so in numerical work. For simulations of classical
statistical models several algorithms~\cite{Machta,Tomita} have been
proposed, that automatically adjust to the critical point. Using ideas
from invasion percolation Machta {\it et al.}~\cite{Machta} proposed an invaded
cluster algorithm where critical clusters are generated
directly, yielding the critical temperature as output.
More recently, Tomita and Okabe~\cite{Tomita} have suggested using a simpler
method called probability-changing cluster algorithm to localize the transition temperature of Ising and Potts models within the
framework of cluster algorithms. In Swendsen and Wang or Wolff algorithms~\cite{SwendsenWolff}, they propose to, {\it during} the simulation, either increase
or decrease the probability $p$ of connecting spins in the same cluster,
and hence the temperature, according to whether the last clusters were
percolating or not.
Thus, systematically moving towards the cluster distribution at the critical point.

The generalization of either algorithm to the simulation of quantum
systems would clearly be highly desirable but is difficult for several
different reasons. Firstly, the structure of the clusters are more
complicated than for classical systems and their relation to percolation
physics is less clear. During the generation of a cluster it is not
possible to assign a probability according to which a single ``spin" is
added to the cluster. Secondly, the dynamical critical exponent
$z$, that links real and imaginary time correlations, is usually
not known~\cite{Z}. Hence, a direct generation of critical clusters would
appear difficult due to the lack of knowledge of the relation between
real-space and imaginary time correlations. A resolution of these
difficulties would be very interesting and here we discuss a first step in
this direction. We propose to generalize the approach of Tomita and
Okabe~\cite{Tomita} to Quantum Monte Carlo (QMC) simulations performed using the loop algorithm~\cite{Evertz1}. We show that adjusting the critical parameter
during the simulation according to whether an estimate of the correlation
length is larger or smaller than some threshold, allows for a very precise
determination of the quantum critical point. With relatively modest effort most
of the associated critical exponents can also be determined. 

The method we propose is closely related to the one suggested by Tomita
and Okabe~\cite{Tomita}. During the simulation a variable measuring the
criticality of the system, in our case the correlation length $\xi$,
is continuously monitored and one then changes
a control parameter, $\lambda$, according whether or not this critical
variable satisfies a criterion.
Here $\lambda$ can be the
temperature $k_BT=1/\beta$ as in~Ref.~\onlinecite{Tomita} or directly a parameter of
the Hamiltonian, controlling the transition. The algorithm should then
equilibrate $\lambda$ to an average critical value $\lambda_c(L)$,
depending on the system size $L$, where
the specified criterion is satisfied as an equality. The most difficult
thing one has to do with this algorithm is to select a
simple numerical criterion indicating whether the critical variable is
in the ordered phase or not.
As discussed in~\cite{Tomita}, by determining the critical value $\lambda_c(L)$
for different linear sizes $L$ of systems, it is possible to obtain
$\lambda_c$ by extrapolating to the thermodynamic limit. 

Let us briefly describe the method for a system of linear size L. 
First, one has to choose an initial value $\lambda_0(L)$ for the parameter
$\lambda$. $\lambda_0(L)$ can be chosen according to a physical intuition of
the critical point, or as the result of a very short preceding simulation at
this size, or of the simulation for a shorter linear size.
In most cases it is useful to choose $\lambda_0(L)$ to be in a phase with
a finite correlation length.
Then a certain number $N_{\rm warm}$ of ``warmup'' Monte Carlo steps are made
to help the system to locate approximately $\lambda_c(L)$. During these
steps, $\lambda$ is modified by a certain amount $\pm \Delta_0 \lambda$ every
$N_t$ sweeps, according to our criterion which we will discuss later. The $N_t$ sweeps are needed to ``thermalize'' the system to
the current value of $\lambda$ to avoid autocorrelation problems (if
$\lambda$ is changed at every step, the configurations might be too
correlated). 
Note that $\Delta_0 \lambda$ is not necessarily fixed,
and can for example be changed linearly from $\Delta_0 \lambda$ to the
smaller fixed value $\Delta \lambda$ used in the ensuing part of the
calculation as discussed in~\cite{Tomita}: this
permits the system to take big steps in $\lambda$ in the early stages of
the simulation if $\lambda_0(L)$ is far from the critical value $\lambda_c(L)$.
When the $N_{\rm warm}$ steps are
finished, one makes $N_{\rm meas}$ steps of measurement
with the same procedure : every $N_t$ steps, $\lambda$ is changed by a small
and fixed value $\pm \Delta \lambda$ and estimates of observables
(including $\lambda$) during the $N_t$ steps are recorded.  At the end of
these $N_{\rm meas}$ sweeps, we get a distribution of $\lambda$ centered
around the critical value, $\lambda_c(L)$, for this system size. For the
interpretation of the final distribution of $\lambda$ it is important to
have used a fixed value of $\Delta\lambda$ during the $N_{\rm meas}$ steps
of measurement. We usually choose $\Delta\lambda \sim \frac{1}{V}$ where $V$
is the volume of the system. Finally, in order to determine $\lambda_c$, the
calculation is repeated for different linear sizes $L$ and an extrapolation
to the thermodynamic limit is performed, in principle allowing for a
determination of the correlation length exponent $\nu$. Other critical
exponents can be determined in this step with the scaling of other observables.

Now we turn to a discussion of the criteria used to lower or raise
$\lambda$. In Refs.~\onlinecite{Machta,Tomita} two criteria based on the
percolation of the clusters were used. These 
criteria, which are referred to as the extension and topological rule in the
related invasion cluster algorithm proposed by 
Machta {\it et al.} \cite{Machta} for classical spin systems, can in principle 
also be used in QMC simulations with the loop algorithm:
percolation or winding of loops can be used to determine the change in
$\lambda$.
However, the interpretation is in this case less clear. Most notably
the exponential distribution of the size of the clusters is very wide and not very useful for
a determination of $\lambda_c$. 
In particular, due to finite-size effects, it is not likely 
to change qualitatively when
$\lambda$ is tuned through the transition. 
Eventually, an algorithm based on
the cluster distribution in the thermodynamic limit at $T=0$ would be
interesting to pursue, but the present method is in our opinion simpler and
more general.
Hence, we choose a different criterion: $\lambda$ is modified according to an
estimate of the correlation length $\xi$ during the $N_t$ steps. It is the same
kind of criterion used in the fixed cluster
algorithm~\cite{Liverpool}. It should be noted that this criterion is not necessarily
related to the loops of the algorithm and can probably be used in other
Monte Carlo simulations which do not admit a cluster algorithm. However,
here the loop algorithm is very helpful because it quickly equilibrates
configurations during the $N_t$ sweeps and because it allows the use of
improved estimators~\cite{Baker} for correlation lengths. Note the importance
of using a reasonably large $N_t$ to have a good estimate of the correlation
length. 

In the following, we will apply and check this method
to quantum $S=1$ spin systems in order to determine the critical value of the
perpendicular coupling, $J^\perp$, necessary for antiferromagnetic order
at T=0 for coupled $S=1$ spin chains.
The $S=1$ spin chain is known to
be disordered~\cite{Haldane} with a gap $\Delta_H\simeq 0.41J$ and a finite
correlation length $\xi \simeq 6$~\cite{White}. On the other
hand, the $S=1$ square antiferromagnet is N\'eel ordered at
$T=0$~\cite{square} and it is known that a finite
inter-chain coupling $J^\perp_c$, at which the system
orders, exists~\cite{Sakai,Kim,Matsumoto}. 
We consider the following Hamiltonian:
\begin{equation}
H=\sum_{i=1}^{L}\sum_{j=1}^{L_y}\left[J{\bf S}_{i,j}\cdot{\bf S}_{i+1,j}+
J^\perp {\bf S}_{i,j}\cdot{\bf S}_{i,j+1}\right],
\end{equation}
with integer $S=1$ spins and periodic boundary conditions.
Previous work on this transition~\cite{Sakai,Kim,Matsumoto}
has estimated the critical coupling and the 
exponents with the most recent estimate~\cite{Matsumoto} 
for $J^\perp_c$ being $J^\perp_c=0.043648(8)$, a value surprisingly small
compared to the Haldane gap $\Delta_H\simeq 0.41J$.

We investigate this quantum phase transition by setting our parameter
$\lambda$ as being the inter-chain coupling $J^\perp$. Our first criterion
(A) is the
following: if the calculated correlation length along the chains is larger
than the system size ($\xi_x>L$), then the system is in an ordered
phase for this value of $J^\perp$ and we decrease $J^\perp$. If $\xi_x<L$,
we increase $J^\perp$. The dividing criterion is thus:
\begin{equation}
{\rm A:}\ \ \xi = L.
\end{equation}
We then record the distribution of $J^\perp$, and estimate
$J^\perp_c(L)$ for each lattice size. This criterion gives a good estimate
of the value of the critical point, but will fail in determining it exactly
and in determining critical exponents. This is due to the fact that an
estimate of $\xi$ for a system size where $\xi\simeq L$ suffers from
pronounced finite-size effects. In particular, the extrapolation of
$J^\perp_c(L)$ to the thermodynamic limit does not follow a simple scaling
law allowing for a determination of the critical exponents. It is a
well-known numerical fact that finite size corrections are significantly
reduced when the lattice size is a few times larger than the calculated
correlation length. Typically, if $6 \xi<L$, finite-size corrections should
be relatively unimportant.
Hence, we use a second criterion (B)
\begin{equation}
{\rm B:}\ \ \xi=\frac{L}{6},
\end{equation}
in order to reduce the finite-size corrections.

We use the continuous time QMC loop algorithm~\cite{Evertz1}
where spins 1 are simulated by two symmetrized spins $1/2$~\cite{Todo}
in the imaginary time direction. In order to study this quantum phase
transition it is crucial to set $T\simeq 0$ and for this purpose we mainly use
the recently proposed improvement of loop algorithm~\cite{Evertz2} which
allows for simulations directly at $\beta =\infty$. Observables are
measured with the help of improved estimators~\cite{Baker}. In particular,
the correlation length $\xi$ is measured with the second moment
method~\cite{Cooper,Baker,Todo}. In general,
simulations have been made on $L$x$L_y$ lattices. For the smaller lattice
sizes, square lattices were used with $T=0$~\cite{Evertz2}; for the largest
lattice sizes $L_y<L$ (with $L_y >> \xi_y$) was used and simulations were
performed at a finite temperature significantly smaller than the smallest
gap in the system. We typically set $N_t=10^3$, $N_{{\rm warm}}=10^5$ and
$N_{{\rm meas}}=10^6$ sweeps.

\begin{figure}
\begin{center}
\psfrag{Jperp}{${\rm J^\perp}$}
\epsfig{file=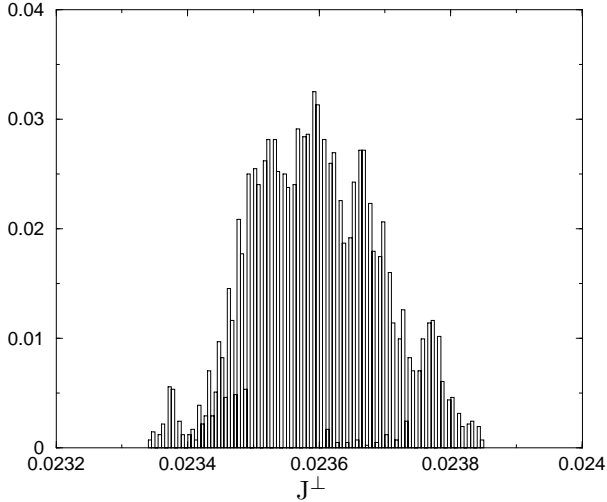,width=8cm}
\caption{Histogram of $J^\perp$ obtained for the $L=66,L_y=10$ lattice at $T=0$
with the criterion B ($\xi=\frac{L}{6}$). The distribution is peaked around
$J^\perp_c(66)=0.0236(1)$.}
\label{fig:histo}
\end{center}
\end{figure}

We now turn to a discussion of our results. In Fig.~\ref{fig:histo} we show
an example of the distribution of $J_{\perp}$ obtained during the simulation
for $L=66$ using criterion B.
%%%
%% FIG 1
%%%
The distribution is sharply peaked around a well defined mean value
allowing for an easy determination of $J^\perp_c(L)$. 
In principle additional information pertaining to the critical exponents should
be extractable from a detailed study of the scaling of the complete
distribution with the system size $L$. For the simulations we have performed
we did not have sufficient statistics to exploit this possibility.
%%%
%% FIG 2
%%%
\begin{figure}
\begin{center}
\psfrag{Jperp}{${\rm J^\perp}$}
\epsfig{file=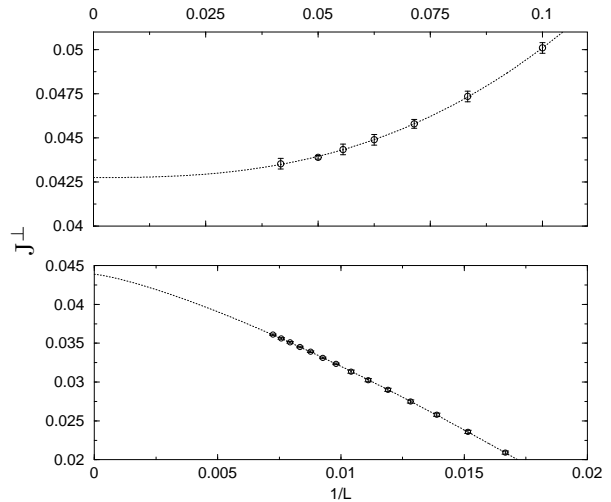,width=8cm}
\caption{$J^\perp_c(L)$ as a function of 1/L for the two different criteria
A (top), B (bottom) used. Lines are fits to the equation~(\ref{pl1}) and
give $J^\perp_c=0.0428(4)$ for the criterion A and $J^\perp_c=0.0438(4)$ for
the B criterion. The values of $J^\perp_c(L)$ found for the first (second)
criterion are always larger (smaller) than the critical coupling.}
\label{fig:Jperp}
\end{center}
\end{figure}
In Fig.~\ref{fig:Jperp} we show $J^{\perp}_c(L)$ as a function of $L$ for the two criteria, A and B. 
As previously mentioned the criterion A suffers from pronounced finite-size
effects. In particular, criterion A always {\it over estimates} the critical
coupling. This might appear counter intuitive since the correlation length
is expected to diverge at $J^\perp_c$. However, due to finite-size effects,
the estimated correlation length never surpasses the linear
size of the system until well inside the antiferromagnetic phase.
One should note that for a finite quantum system there is always a finite
gap restraining the correlation length.
In order to extract the critical exponent we can as a first approximation
fit the results obtained to a power law:
\begin{equation}
J^\perp_c(L)=J^\perp_c + A L^{-a}.
\label{pl1}
\end{equation}
Here $J^\perp_c$ is the critical coupling (in the thermodynamic limit) and
$A$ a constant. For the criterion B ($\xi_x=L/6$), $a=1/\nu$ where $\nu$ is
the correlation length exponent. For the criterion A, $a \neq 1/\nu$
due to finite-size corrections and the $L$ dependence is non-trivial.
We obtain $J^\perp_c=0.0438(4)$ and $\nu=0.77(2)$ for the criterion 
B, and
$J^\perp_c=0.0428(4)$ for the criterion A. The more reliable
values are of course the first ones, but one can note that even with the
criterion A, we have a good estimate of $J^\perp_c$. We can
try to improve on this estimate by including corrections to scaling. In
analogy with previous work on finite-size scaling~\cite{fscaling} we include
a first order finite-size correction to the correlation length.
\begin{eqnarray}
\frac{L}{6} & = & \xi - be^{-L/\xi}\nonumber\\
\xi & = & a(J^\perp_c-J^\perp_c(L))^{-\nu}.
\label{corr}
\end{eqnarray}
Including this finite size correction we find $\nu = 0.72(2)$, 
$J^\perp_c=0.0436(2)$.  In essence this correction amounts to a relatively
precise estimate of the correlation length $\xi$ in the thermodynamic limit
for the given value of $J^\perp_c(L)$. This is then fitted to the expected
form for the divergence of the correlation length as $J^\perp_c$ is
approached. In principle higher order corrections could be included, we
have verified that in the present case they are extremely small.
This value of $J^\perp_c$ is in agreement with the recent QMC
``traditional'' simulations~\cite{Matsumoto} and is more precise than
older results obtained with exact diagonalizations~\cite{Sakai} or
QMC~\cite{Kim} calculations.

\begin{figure}
\begin{center}
\epsfig{file=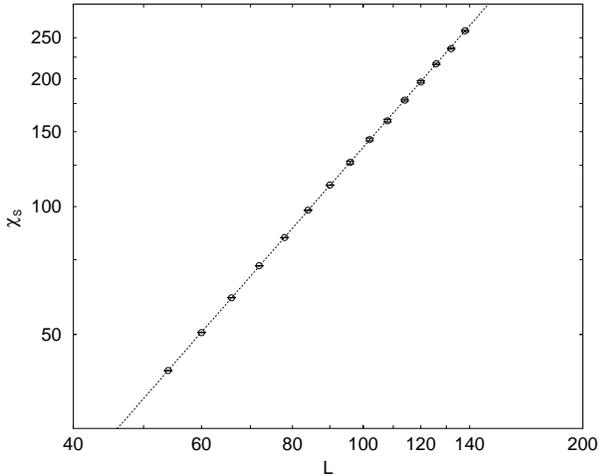,width=8cm}
\caption{The staggered susceptibility $\chi_s$ as a function of $L$ on a
log-log scale. The power-law fit~(\ref{pl2}) displayed as the line yields
$\eta=0.038(3)$.}
\label{fig:Chis}
\end{center}
\end{figure}

Once a reliable estimate of $J^\perp_c(L)$ is obtained high-precision
estimates of additional critical exponents can be obtained by performing
calculations directly at $J^\perp_c(L)$. In Fig.~\ref{fig:Chis} we show results
%%%
%% FIG 3
%%%

of the size dependence (on log-log scale) of the staggered susceptibility
$\chi_s$ with the criterion B ($\xi_x=L/6$). According to standard
finite-size scaling theory this quantity should scale as
\begin{equation}
\chi_s(L) \sim L^{2-\eta},
\label{pl2}
\end{equation}
even with $J^\perp_c(L)$ a function of $L$. Additional finite size
corrections of the form~(\ref{corr}) for $\chi_s$ were not included.
Fitting to this form~(\ref{pl2}), we obtain $\eta=0.038(3)$ in complete
agreement with previous estimates. 
The obtained values of $\eta$ and $\nu$ for the critical exponents are in
agreement with Matsumoto {\it et al.}'s results~\cite{Matsumoto}.
The exponents are very close 
to the values for the 3D classical Heisenberg model~\cite{MCC} confirming the
expectation that two dimensional quantum phases transitions of quantum
Heisenberg magnets belong to the university class of classical
magnets~\cite{Chubukov} as seen in in other
simulations~\cite{Troyer,Matsumoto}.

%\section{Outlines and conclusions}\label{conclusion}

In this paper, we have proposed a very simple method to determine critical
points in QMC simulations as a generalization of the probability-changing
cluster algorithm~\cite{Tomita} and applied it to determine
the critical value of inter-chain coupling for spin $S=1$ chains and calculate
critical exponents for this quantum phase transition. The results are in 
agreement with previous estimates.
It is important to underline that this method is easy to employ and
adapt to other phase transitions and that it can save a lot of computational
effort. In particular, no a priori knowledge of $z$ is needed.
Most other methods, as for instance the use of Binder
cumulants~\cite{Binder}, would require the estimation of a critical quantity
for many different system sizes for each value of the control
parameter. The present method only requires the knowledge of the critical
quantity for one system size for each of the control parameter. 
As mentioned in the introduction, it would be highly desirable to extend
also the invaded cluster algorithm~\cite{Machta} to the simulation of
quantum systems and we hope the present work could be a first step in that direction.

\acknowledgements
We would like to thank S. Todo for valuable discussions and comments on the
manuscript.

\end{document}